\documentclass[showpacs,preprintnumbers,amsmath,amssymb,aps,prb,floatfix,reprint,superscriptaddress]{revtex4-1}

\usepackage{graphicx}
\usepackage{hyperref}
\usepackage{amsmath}

\newcommand{\der}[2]{\frac{\mathrm{d}#1}{\mathrm{d}#2}}

\newcommand{\pder}[2]{\frac{\partial#1}{\partial#2}}

\begin{document}
\title{Effects of spatially engineered Dzyaloshinskii-Moriya interaction in ferromagnetic films}
\author{Jeroen Mulkers}
\email[Email: ]{jeroen.mulkers@uantwerpen.be}
\affiliation{Departement Fysica, Universiteit Antwerpen, Groenenborgerlaan 171, B-2020 Antwerpen, Belgium}
\affiliation{DyNaMat Lab, Department of Solid State Sciences, Ghent University, Ghent, Belgium}
\author{Bartel Van Waeyenberge}
\affiliation{DyNaMat Lab, Department of Solid State Sciences, Ghent University, Ghent, Belgium}
\author{Milorad V. Milo\v{s}evi\'c}
\affiliation{Departement Fysica, Universiteit Antwerpen, Groenenborgerlaan 171, B-2020 Antwerpen, Belgium}

\date{\today}

\begin{abstract}
The Dzyaloshinskii-Moriya interaction (DMI) is a chiral interaction that favors formation of domain walls. Recent experiments and \textit{ab initio} calculations show that there are multiple ways to modify the strength of the interfacially induced DMI in thin ferromagnetic films with perpendicular magnetic anisotropy. In this paper we reveal theoretically the effects of spatially varied DMI on the magnetic state in thin films. In such heterochiral 2D structures we report several emergent phenomena, ranging from the equilibrium spin canting at the interface between regions with different DMI, over particularly strong confinement of domain walls and skyrmions within high-DMI tracks, to advanced applications such as domain tailoring nearly at will, design of magnonic waveguides, and much improved skyrmion racetrack memory.
\end{abstract}


\maketitle

\section{introduction}

The ground state magnetization of a thin ferromagnetic film with perpendicular magnetic anisotropy (PMA) is uniform at nanoscale. Only at larger length scales, domains of `up' and `down' magnetization can be stabilized by the long range dipolar interactions, a process called demagnetization. However, in ultrathin heterostructures where a ferromagnetic layer, e.g., a Co layer, is coupled to a nonmagnetic layer with a strong spin-orbit coupling, e.g., a heavy metal like Pt, one has to take into account the interfacially induced Dzyaloshinskii-Moriya interaction (DMI).\cite{Dzyaloshinsky1958,Moriya1960,Dzyaloshinsky1964,Crepieux1998} This chiral interaction favors rotation of the magnetization at short length scales, and, when strong enough, stabilizes chiral spin structures such as cycloids (parallel N\'eel walls) and skyrmions (closed N\'eel walls).\cite{Bogdanov1994,Bogdanov2001,Ezawa2011,Kiselev2011}

The confining effect of the boundaries and the shape of the ferromagnetic film on the chiral spin structures has already been thoroughly studied.\cite{Rohart2013,Chui2015,Mulkers2016,Navau2016} This confining effect is utilized in the design of skyrmion-based devices, e.g., a skyrmion racetrack memory in which skyrmions can be moved by spin currents.\cite{Fert2013,tomasello2014,Zhang2015,Iwasaki2014}  Another way to control the behavior of chiral spin structures in these devices, is by locally applying an electric field. Such external electric field changes the magnetic anisotropy and consequently changes locally the relative strength of the different types of magnetic interactions.\cite{Zhang2015b,Kang2016,Fook2016} In this paper, we suggest a third ingredient to design advanced skyrmion-based devices: \emph{the spatially engineered DMI}.

The effective DMI strength depends on the thickness of the ferromagnetic layer and the stacking of the ferromagnetic layer and heavy-metal layers.\cite{Hrabec2014,Kim2015,Cho2015,Belmeguenai2016,Yang,Tacchi2016,Ma2016} In principle, one can alter the DMI strength locally by changing the thickness of the ferromagnetic layer. This, however, is very challenging to realize experimentally. Furthermore, this causes nontrivial (3D) boundary effects in the ferromagnetic film. Instead, we suggest to use a uniform, extended ferromagnetic layer and alter the DMI strength locally by using lithographic techniques to (partially) change or remove the heavy metal layer on top of the ferromagnet.\cite{Balk2016,Wells2017} Modifying the DMI, by altering the covering heavy metal layer, can possibly change other material parameters, such as the magnetic anisotropy. In our theoretical study, we want to identify the exclusive effect of spatially varied DMI and thus keep the other material parameters homogeneous. In the considered heterochiral films, we show that a spatially engineered DMI gives rise to a plethora of unique effects. For example, the uniform state shows spin canting at the interface between regions with different DMI; high-DMI regions strongly confine chiral spin structures, such as single domain walls, cycloids, and skyrmions; DMI engineering can be used to design tracks for magnons and skyrmions, with improved characteristics compared to other existing realizations.

The paper is organized as follows. In Sec.~\ref{sec:methods} we briefly recapitulate the micromagnetic framework of ferromagnetic films with an interfacially induced DMI. In Sec.~\ref{sec:statics}, we discuss characteristic static magnetization configurations in ferromagnetic films with a spatially engineered DMI. This includes canting of the magnetization at interfaces where DMI changes, confined cycloids in high-DMI strips, and confined skyrmions in high-DMI disks and strips. Some possible applications of a spatially engineered DMI are discussed in Sec.~\ref{sec:applications}, ranging from manipulation of size and shape of confined domains, relevant to miniaturization and reliability of magnetic memory, over spin waveguides,\cite{Xing2016,Garcia-sanchez2014,Garcia-Sanchez2015,Borys2016} to an advanced design of skyrmion racetrack. Our results are summarized in Sec.~\ref{sec:conclusion}.

\section{\label{sec:methods}Micromagnetic framework}

We describe the magnetization of a ferromagnetic film by a 2D field~$\vec{M}(x,y)=M_{sat}\vec{m}(x,y)$ with magnetization modulus~$|\vec{M}|=M_{sat}$ and magnetization direction~$\vec{m}(x,y)$. The local free energy density, related to the magnetization~$\vec{M}$, has multiple sources: exchange, anisotropy, Zeeman interaction, DMI, and demagnetization. We approximate the demagnetization energy by using an effective anisotropy~$K_{\mathrm{eff}}=K-1/2\mu_0 M^2_{\mathrm{sat}}$.\cite{Coey} By doing this we neglect the volume magnetic charge contribution of the N\'eel domain walls. This is justified since we will confine domain walls in regions with a strong DMI, where the effect of the volume magnetic charges on the energy is insignificant compared to the negative DMI energy of N\'eel domain walls.\cite{Thiaville2012}
The expressions for the remaining energy-density terms are, respectively,
\begin{eqnarray}
    \varepsilon_{\mathrm{ex}} &=& A \textstyle{\left[ \left(\pder{\vec{m}}{x}\right)^2 + \left(\pder{\vec{m}}{y}\right)^2\right]}, \\
    \varepsilon_{\mathrm{ext}} &=& -\vec{B} \cdot \vec{m} M_{\mathrm{sat}},\\
    \varepsilon_{\mathrm{anis}} &=& K_{\mathrm{eff}}(1-m_z^2), \\
    \varepsilon_{\mathrm{dmi}} &=& D \textstyle{\left[ m_x\pder{m_z}{x} - m_z\pder{m_x}{x}+m_y\pder{m_z}{y}-m_z\pder{m_y}{y} \right]}, \label{eq:dmidens}
\end{eqnarray}
with exchange stiffness~$A$, DMI strength~$D$, anisotropy constant~$K_{\mathrm{eff}}$, and external magnetic field~$\vec{B}$. To simplify the notation, we introduce the exchange length~$\xi=\sqrt{A/K_{\mathrm{eff}}}$ and the critical DMI strength~$D_c =4\sqrt{AK_{\mathrm{eff}}}/\pi$. Note that the energy of a single domain wall~$E_{\mathrm{wall}}=4\sqrt{AK_{\mathrm{eff}}}-\pi D$ is positive for DMI strengths below $D_c$ and negative otherwise.~\cite{Rohart2013}

The energy expression can be simplified for 1D problems for which we make the assumption that the magnetization varies only along the $x$ direction and the magnetization has no $y$ component (since this lowers the DMI energy).\cite{Rohart2013} The magnetization is then fully defined by its angle~$\theta$ with respect to the $z$ axis: $m=(\sin\theta,0,\cos\theta)$. Under these conditions, the local free energy density becomes
\begin{equation}
    \varepsilon = A\left(\der{\theta}{x}\right)^2 - D\der{\theta}{x} + K_{\mathrm{eff}} \sin^2\theta. \label{eq:1d}
\end{equation}

The dynamics of the magnetization is governed by the Landau-Lifshitz-Gilbert (LLG) equation
\begin{equation}
    \vec{m}_t = \frac{\gamma_{\mathrm{LL}}}{1+\alpha^2} \left( \vec{m} \times \vec{H}_{\mathrm{eff}} + \alpha \left[ \vec{m} \times( \vec{m} \times \vec{H}_{\mathrm{eff}}) \right] \right),\label{eq:llg}
\end{equation}
with gyromagnetic ratio $\gamma_{\mathrm{LL}}$ and damping factor $\alpha$. The effective magnetic field is the functional derivative of the magnetic free energy $E=\int\varepsilon \mathrm{d} V$ with respect to the magnetization: $\vec{H}_{\mathrm{eff}}=-\delta E/\delta\vec{m}$.

We will consider the magnetization to be a continuous field. If we do not impose this property, the exchange energy will be infinite at the points of discontinuity, rendering the magnetization unable to capture the physics of the underlying atomic magnetic moments. In contrast to the continuity of the magnetization field, there is no reason to impose continuity on its spatial derivatives. It is sufficient to have a semi differentiable field, meaning that the field is continuous and that at every point one can calculate the left and right derivative. This requirement makes the derivatives continuous \emph{almost everywhere} (except at material interfaces), allowing us to compute definite integral functionals depending on the magnetization and its first derivatives, in particular the energy functional and the effective field.


With these requirements in mind, we can calculate the DMI energy of magnetization configurations in films with regions of different DMI. The DMI energy density~$\varepsilon_{\mathrm{dmi}}$ can be integrated by considering a different left and right derivative at the interface where DMI changes, multiplied by the corresponding DMI strength. Although this is not appropriate for a continuously varying DMI, it is fine for any finite discretization (where a continuous function is represented as a stepped series). Moreover, at the atomic scale the DMI strength is defined between two distinct magnetic moments, thus assumes steplike change where material properties change.

In most cases it is hard, or even impossible, to minimize the energy or solve the LLG equation analytically, which makes numerical computations inevitable. For one-dimensional problems, in which the magnetization can be characterized by a single angle $\theta(x)$, we minimize the energy functional by discretizing the magnetization on a fine spatial grid~$\theta_i = \theta(i\Delta x)$ with $\Delta x \ll \xi$, and minimizing the total energy $E(\theta_i)$ in which the spatial derivatives are approximated by finite differences. For the more challenging numerical computations and simulations, we use the finite-differences-based micromagnetic simulation package Mumax3~\cite{Vansteenkiste2014}, where we incorporated inhomogeneous DMI in the above described fashion (this feature is made publicly available in Mumax version 3.9.3 and later). The cell size in such simulations is set to $(0.1,0.1,0.1)\xi$ and demagnetization is approximated with an effective anisotropy.

In both numerical approaches, the derivative of the magnetization in a cell is approximated by the sum of the left and right first-order finite differences. This allows us to study the effect of a regionally different DMI with an interface running through the centers of adjacent cells. Note that for uniform material parameters, the sum of left and right finite differences yields the second-order central difference. Only at the interface cells, we end up with a first-order approximation of the magnetization derivatives. It is worth mentioning that using the sum of left and right first-order finite differences turns out to be equivalent with an atomistic spin model on an orthorhombic lattice with lattice parameters the same as the cell dimensions in the finite-difference micromagnetic model (see Appendix). These cell dimensions are usually larger than the typical distance between atoms, but as long as the magnetization varies slowly, the micromagnetic model and the spin model on the atomistic scale will yield approximately the same magnetization density.

\section{\label{sec:statics}Magnetostatics of heterochiral films}

\subsection{\label{sec:uniform}Quasi-uniform state}

\begin{figure}[b]
\centering
\includegraphics{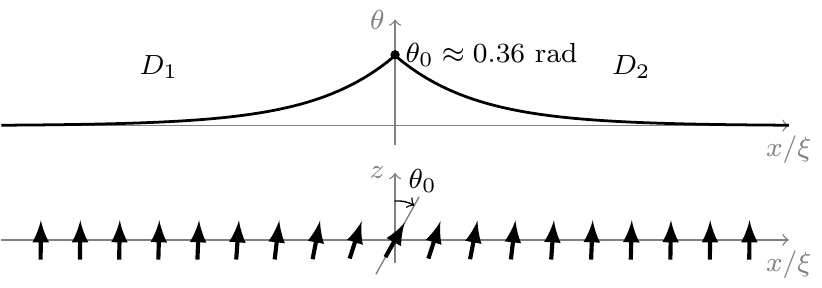} 
\caption{Canting of the spins in the quasiuniform state of a ferromagnetic film, at the interface of two regions with different DMI strengths ($D_1-D_2 = 1.1 D_c$).}
\label{fig:canting}
\end{figure}

As the simplest case of a heterochiral film, we first consider a 1D model with DMI strength~$D_1$ on the left $(x<0)$ and DMI strength~$D_2<D_1$ on the right $(x>0)$, as depicted in Fig.~\ref{fig:canting}. After relaxing the uniform magnetized state, significant canting of the magnetization is observed at the interface between regions with different DMI (in the present case, at $x=0$). We denote the canting angle of magnetization at the interface as $\theta_0$.

After minimizing the energy functional (\ref{eq:1d}) with the Euler-Lagrange method, we obtain the expression for the magnetization
\begin{equation}
    \theta(x) = 2\arctan\left(\mathrm{e}^{-\left|x/\xi\right|}\tan\frac{\theta_0}{2}\right),
\end{equation}
which bears similarities with the expression for a domain wall. Note that one still has to determine the angle~$\theta_0$. In order to do this, we calculate the energy difference with respect to the energy of the uniform state:
\begin{eqnarray}\label{eq:E_uni}
    E^{\mathrm{1D}}_{\mathrm{cant}} &=& \int \varepsilon \mathrm{d}x \nonumber \\
                                    &=& 4\sqrt{AK_{\mathrm{eff}}}(1-\cos\theta_0)+(D_2-D_1)\theta_0.
\end{eqnarray}
The first term is the increase in exchange and anisotropy energy due to the canting with angle~$\theta_0$ at the interface. The second term is the DMI energy which depends on the difference between DMI strengths $D_1$ and $D_2$.  Minimizing the energy difference, by varying the angle~$\theta_0$, yields
\begin{equation}
    \theta_0 = \arcsin \frac{D_1-D_2}{4\sqrt{AK_{\mathrm{eff}}}}.
\end{equation}
The canting of the magnetization at the interface depends on the difference between DMI strengths $D_1$ and $D_2$. Even for DMI strengths below $D_c$, canting will occur. Increasing the exchange or the anisotropy will lower the canting of the magnetization~$\theta_0$, i.e., would tend to make the state more uniform. Since the DMI can be positive or negative, the largest canting angle is obtained when $D_1=-D_c$ and $D_2=D_c$, and is equal to $\arcsin(2/\pi) \approx 0.69$ rad.

\subsection{\label{sec:cycloid}Confined cycloids}

\begin{figure}[t]
    \includegraphics{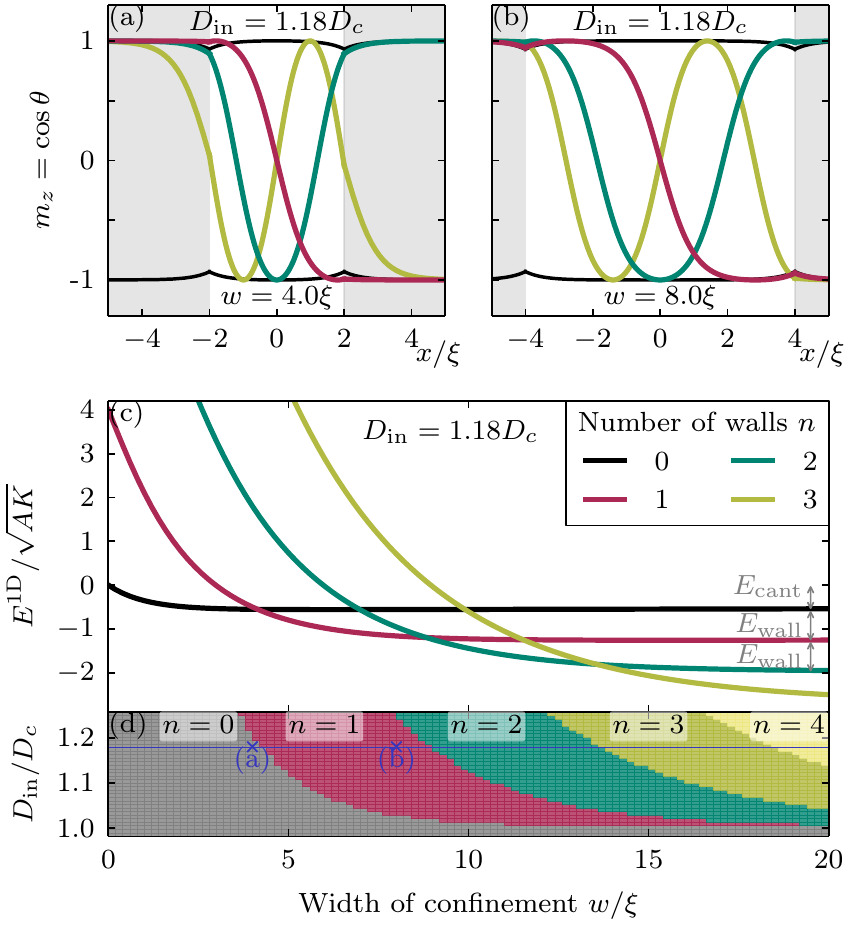} 
    \caption{(a),(b) Magnetization profiles of confined cycloids in high-DMI stripes within a ferromagnetic film, for stripe width~$w=4\xi$ and $w=8\xi$ and DMI strength~$D_{\mathrm{in}}=1.18D_c$. (c) The energies of the cycloids in a function of confinement width~$w$. (d) The ground-state phase diagram of confined cycloids as a function of both width $w$ and DMI strength~$D_{\mathrm{in}}$.}
    \label{fig:cycloid}

\end{figure}

The magnetic ground state of a chiral ferromagnetic film is cycloidal in the case of a strong DMI $|D|>D_c$ and uniform otherwise. Therefore, in a heterochiral film, it is possible to confine a cycloid in a high-DMI region. To illustrate this, we consider a high-DMI strip in the film, surrounded by an extended region without DMI. The energy of the confined cycloid is the sum of the energy in the outer regions, which we calculated analytically in the previous section [Eq. (\ref{eq:E_uni})], and the energy inside the high-DMI strip. We can rewrite the 1D energy functional for this case as
\begin{align}
    E^{\mathrm{1D}} = & \int_{-a}^{+a}  \left[ A\left(\der{\theta}{x}\right)^2-D_{\mathrm{in}}\der{\theta}{x} + K_{\mathrm{eff}} \sin^2 \theta \right] \mathrm{d}x \nonumber \\
                      & + \sqrt{AK_{\mathrm{eff}}} \left(4 -2|\cos\theta_{-a}| -2|\cos\theta_{+a}|\right),
\end{align}
for a cycloid confined in a high-DMI strip with DMI strength~$D_{\mathrm{in}}$ and width~$w=2a$, centered at $x=0$. We minimize the energy~$E^{\mathrm{1D}}$ using a conjugate gradient method. This yields a stable magnetization~$\theta(x)$, including the magnetization at the interfaces~$\theta_{\pm a}$. The stabilized number of domain walls in the confined cycloid depends on the initial guess of the magnetization. The energy and magnetization profiles of the confined cycloids, corresponding with the lowest energy states, are shown in Fig.~\ref{fig:cycloid}.

The spin canting at the interface, discussed in the previous section, is clearly visible for the quasiuniform state.  This spin canting also occurs when the number of domain walls is small in a wide high-DMI strip [see single-wall case in Fig.~\ref{fig:cycloid}(b)]. The energy of the cycloidal state with $n$ domain walls converges to the sum of wall energies~$E_{\mathrm{wall}}$ and the energy of canted spins at the interface ~$E_{\mathrm{cant}}$, i.e., $n E_{\mathrm{wall}} + E_{\mathrm{cant}}$ for large confinement widths~$w$. In other words, adding walls lowers the energy for large~$w$. On the other hand, a large number of domain walls is not preferred in narrow confinement regions due to the resulting high exchange energy. These competing effects are notable in the phase diagram of the ground state, shown in Fig.~\ref{fig:cycloid}(d). The ground state is quasiuniform for DMI strength~$D_{\mathrm{in}}$ below $D_c$, or for small~$w$. The number of domain walls in the ground state increases for increasing DMI strength or an increasing confinement width~$w$ of the high-DMI strip.

This phase diagram is similar to the phase diagram of cycloids confined in finite ferromagnetic films with homogeneous DMI~\cite{Mulkers2016}. However, there is a fundamental difference between the two design methods regarding the stability of domains. The energy barrier related to the annihilation of domain wall at the edge of a monochiral ferromagnet is smaller than the energy needed to collapse an entire domain in a high-DMI strip within a low-DMI film.
A Bloch point is unavoidable when collapsing an entire domain in an extended film. This leads to a very high energy barrier (infinite exchange energy in the continuum approximation). Pushing a domain out of a film can be done in a continuous manner without the formation of a Bloch point, corresponding with a finite energy barrier. In an atomistic model, the energy barrier to annihilate a skyrmion becomes finite, but it is reasonable to assume that this barrier is still much higher than the energy barrier for the escape of a skyrmion through sample boundary.

\subsection{\label{sec:circskyrmion}Confined skyrmion}

\begin{figure}[t]
    \centering
    \includegraphics{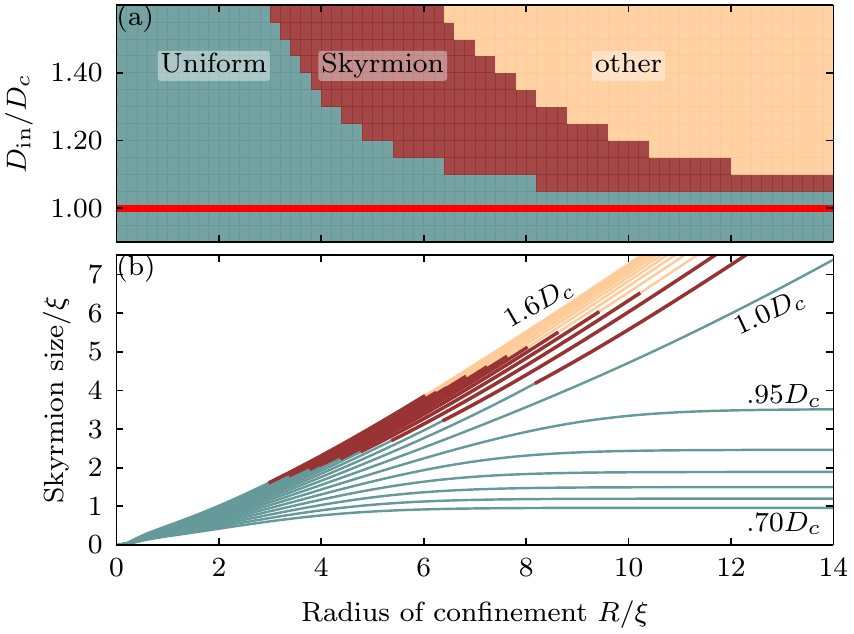} 
    \caption{(a) Ground-state magnetic phase diagram of a high-DMI circular region (with radius~$R$ and DMI strength~$D_{\mathrm{in}}$) within an extended ferromagnetic film without DMI. (b) Skyrmion size as a function of the confinement radius~$R$, for different DMI strength~$D_{\mathrm{in}}$. The red segments indicate a skyrmion as the ground state of the system.}
    \label{fig:circconfskyrmion}
\end{figure}

In the next consideration, we assume that the magnetization~$\theta(r)$ has cylindrical symmetry and the direction of the magnetization is radial, as done previously in Ref.~\onlinecite{Rohart2013}. In that case, the energy functional in polar coordinates becomes
\begin{align}
    E^{\mathrm{2D}}[\theta(r)]  =  2\pi  \int_{0}^{+\infty} \left[ A \left(\der{\theta}{r}\right)^2 + A\frac{\sin^2\theta}{r^2} \right. \nonumber \\
    \left. -D(r) \left(\der{\theta}{r} + \frac{\cos\theta\sin\theta}{r} \right) + K_{\mathrm{eff}}\sin^2 \theta \right] r\mathrm{d}r, \label{eq:circ}
\end{align}
with allowed radially dependent DMI strength~$D(r)$. Here, we consider a strong DMI in a central circular region of radius~$R$, surrounded by an extended region without DMI [$D(r)=D_{\mathrm{in}}\Theta(r-R)$, with $\Theta(r)$ the Heaviside step function]. We then relax the quasiuniform state (no walls to begin with), a confined skyrmion (single closed wall), and a ring domain (two concentric closed walls) by minimizing the energy numerically. Comparing the energies of the three configurations yields the phase diagram of the ground state shown in Fig.~\ref{fig:circconfskyrmion}(a). The quasiuniform state is the ground state for DMI strengths below $D_c$ or for strong confinement (small $R$). For DMI strengths above $D_c$, there is a range of confinement size~$R$ for which the magnetization with a skyrmion in the high-DMI region is the ground state. For a strong DMI and a loose confinement (large $R$), the circular domain is the lowest energy state of the three configurations considered. For larger $R$, one finds higher order ring domains or other cycloidal-like domains (e.g.\ S-shaped ones, beyond our cylindrical approximation), within the parametric area labeled `other' in Fig.~\ref{fig:circconfskyrmion}(a).

The radius of the confined skyrmion, after minimization of the energy functional~(\ref{eq:circ}), is shown in Fig.~\ref{fig:circconfskyrmion}(b). The figure shows clearly the effect of the confinement: The stronger the confinement, the smaller the skyrmion radius. In the continuum approximation, there is no limit on how small one can confine a skyrmion. In real samples, however, substantial shrinking of a skyrmion makes it increasingly unstable, and eventually the skyrmion will collapse.\cite{Rohart2016,Lobanov2016}

\subsection{\label{sec:track}Skyrmion on a track}

\begin{figure}[t]
    \centering
    \includegraphics{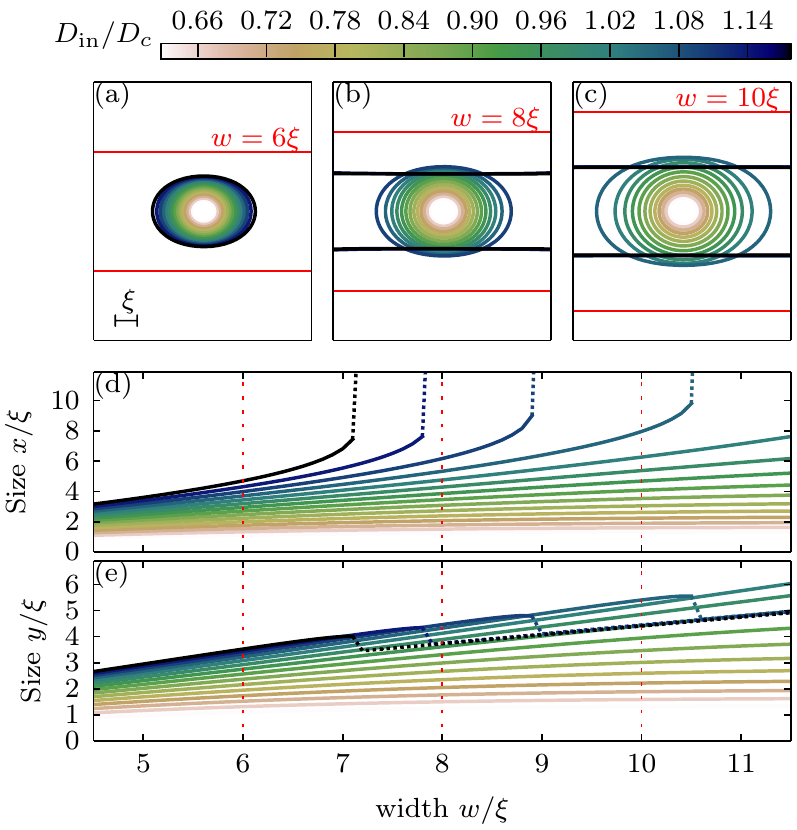} 
    \caption{Isomagnetization outline ($m_z=0$) of a skyrmion in a skyrmion track of width 6$\xi$ (a), 8$\xi$ (b), and 10$\xi$ (c), defined by a stripe with DMI strength~$D_{\mathrm{in}}$ within a larger ferromagnetic film without DMI. Panels (d) and (e) show the size of the skyrmion along the track ($x$ direction) and across the track ($y$ direction). The dashed curves indicate the expansion of the skyrmion into a linear domain.}
    \label{fig:skyrmiondeform}
\end{figure}

In this section, we examine the confinement effects and resulting deformation of a skyrmion within a high-DMI track. A single skyrmion was placed at the center of the track and subsequently relaxed, using the minimizer in Mumax3, for different widths $w$ and DMI strengths~$D_{\mathrm{in}}$ of the track (outside the track, DMI was held at zero). The obtained change of geometry and size of the relaxed skyrmion is shown in Fig.~\ref{fig:skyrmiondeform}.

For low DMI strength~($|D_{\mathrm{in}}|<D_c$), the effect of the confinement is only visible for narrow tracks. The size of the skyrmion converges to its expected size in an infinite film with given DMI, when increasing the width of the high-DMI track. For a stronger DMI in the track, a skyrmion will be elongated in the direction of the track. If the width of the track is above a threshold value, the skyrmion will expand along the track and convert into a stripe domain. This transition is represented in Fig.~\ref{fig:skyrmiondeform}(d) by the `divergent' size of the skyrmion in the $x$ direction.

\begin{figure}[t]
    \centering
    \includegraphics{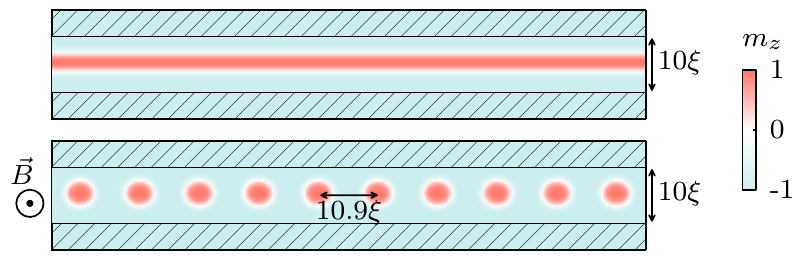} 
    \caption{(a) The cycloidal ground state in absence of an external field, and (b) a skyrmionic ground state for an external field $B=0.3 K_{\mathrm{eff}}/M_{\mathrm{sat}}$ in a high-DMI strip of width $w=10\xi$ and DMI strength $D_{\mathrm{in}}=1.3D_c$. In the shaded area outside the high-DMI strip there is no chiral interaction.}
    \label{fig:skyrmionband}
\end{figure}

In an extended ferromagnetic film with a strong DMI ($|D|>D_c$), the ground state is cycloidal. However, it is possible that the ground state becomes a triangular skyrmion lattice when applying an external field.\cite{Muhlbauer2009} For a high-DMI strip, we see a similar phenomenon. For example, the ground state in a high-DMI track of width~$10\xi$ and $D_{\mathrm{in}}=1.3D_c$ is a single stripe domain. In the presence of an external magnetic field $B=0.3 K_{\mathrm{eff}}/M_{\mathrm{sat}}$, the ground state becomes a skyrmion chain confined in the center of the track (see Fig.~\ref{fig:skyrmionband}). For larger width of the track, the zigzag instability of the skyrmionic chain is expected, in analogy to similar studies on quasi-1D colloidal\cite{Piacente2010} and superconducting vortex systems\cite{Karapetrov2009}.

\section{\label{sec:applications}Exemplified applications}

\subsection{\label{sec:design}Domain design by DMI engineering}

\begin{figure}[t]
    \centering
    \includegraphics{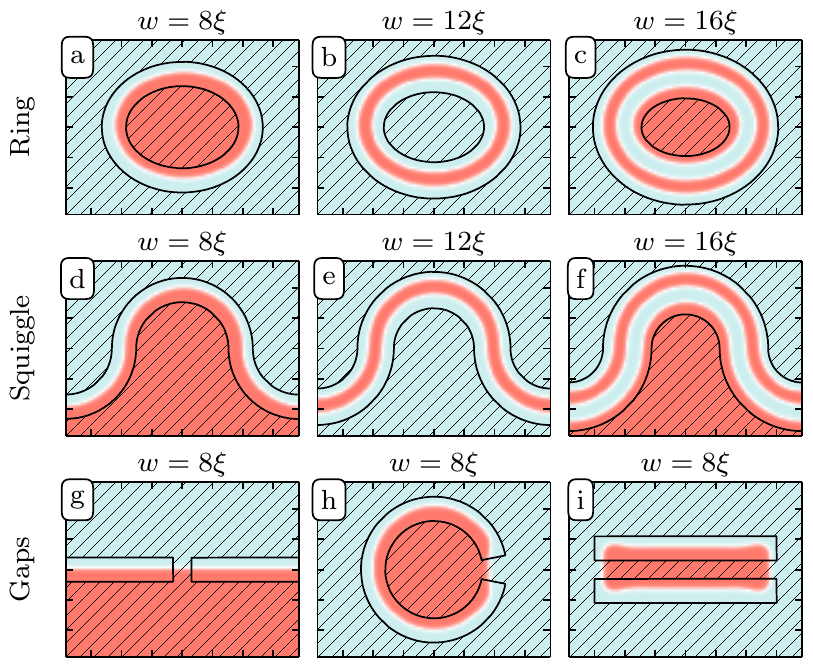} 
    \caption{Spatial magnetization profiles of the lowest energy states for high-DMI strips of different shapes and width~$w$. The DMI strength is $1.18D_c$ inside the strips (outlined by black lines) and zero elsewhere (shaded area). }
    \label{fig:design}
\end{figure}

Considering a ferromagnetic strip with a spatially inhomogeneous DMI strength, a rotation in its magnetization is more favorable in regions with a strong DMI. We already showed that cycloidal states, including a single N\'eel domain wall, can be very effectively confined in straight high-DMI strips. We find that this is also the case for curved high-DMI strips, which means that a spatially-engineered DMI can be used to fix the location of the contained domain wall(s) and thereby design domains of arbitrary shape and size. Figure~\ref{fig:design} demonstrates the proof of principle of such domain design approach by showing the lowest energy states of ferromagnetic films with (curved) high-DMI strips of different widths~$w$ and shape.

Figures~\ref{fig:design}(g)-\ref{fig:design}(i) also demonstrate that small gaps in the high-DMI strips do not affect the end result. This is important to prove that the suggested method to fix the shape and the size of domains is robust against sample imperfections, which may be crucial for experimental realization. Note that the number of domain walls in the lowest-energy state depends on the width~$w$ of the high-DMI region. This dependence is already discussed in detail for straight high-DMI strips in Sec.~\ref{sec:cycloid}. Recall also that in Sec.~\ref{sec:track} we showed that, in a high-DMI strip with $|D_{\mathrm{in}}|>D_c$, a skyrmion expands and eventually forms a cycloidal state with two walls parallel to the borders of the strip. This expansion can be used to create worm domains of desired shape in curved, pre-engineered high-DMI strips [see e.g.\ Fig.~\ref{fig:design}(e)] by first nucleating a skyrmion anywhere within the strip.

\subsection{\label{waveguide} High-DMI waveguides}

\begin{figure}[b!]
    \centering
    \includegraphics{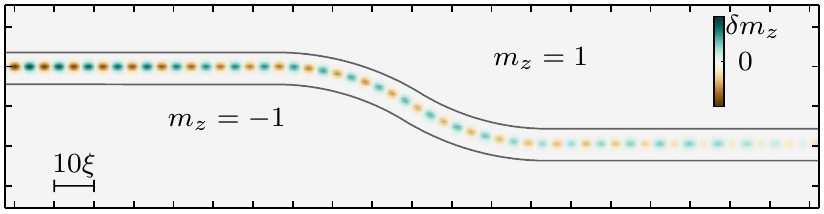} 
    \caption{A spin wave, with angular frequency~$w=0.4 \gamma_{\mathrm{LL}} K/M_{\mathrm{sat}}$, traveling (from left to right) along a single domain wall confined in a curved high-DMI strip. The DMI strength is $D_{\mathrm{in}}=1.18D_c$ inside the strip and zero elsewhere. A damping factor~$\alpha=0.02$ is used. The colormap depicts the dynamic variation of the $z$ component of the magnetization.}
    \label{fig:waveguide}
\end{figure}

Domain walls are known to act as spin waveguides.\cite{Xing2016,Garcia-sanchez2014,Garcia-Sanchez2015,Borys2016} As shown in the preceding subsection, it is possible to fix the position, shape, and number of domain walls using a smoothly curved high-DMI strip and thereby also engineer the path for guidance of the spin waves. An example of a curved spin waveguide, based on a single N\'eel domain wall confined in a curved high-DMI strip, is shown in Fig.~\ref{fig:waveguide}.

\begin{figure}[t!]
    \centering
    \includegraphics{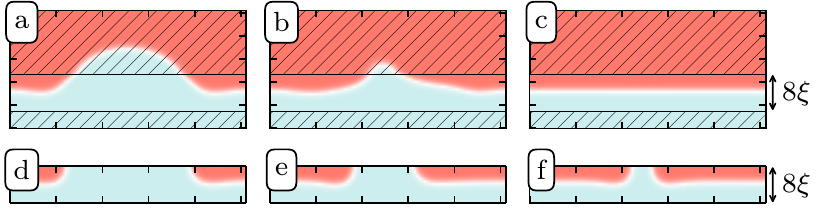} 
    \caption{(a)-(c) Time resolved self recovery of a domain-wall waveguide in a high-DMI strip within an extended, heterochiral ferromagnet, with DMI strength~$D_{in}=1.18D_c$ inside the strip and no DMI elsewhere (shaded area). (d)-(f) Relaxation of a broken domain wall waveguide in a finite, monochiral ferromagnet with DMI strength~$D=1.18D_c$. The initial configurations (a) and (d) are relaxed using the LLG equation with damping parameter~$\alpha=1$.}
    \label{fig:selfrecovery}
\end{figure}

Using a confined domain wall in a high-DMI strip of an extended, heterochiral ferromagnet instead of a domain wall confined by the boundaries of a finite, monochiral ferromagnet, is beneficial for the robustness of the waveguide against deformations of the domain wall guiding the wave. For example, consider a domain wall mostly confined in a high-DMI strip but with a meandering deformation that places the domain wall partially inside the low-DMI region around the waveguide. This deformed domain wall will then relax towards its equilibrium state, back in the center of the confining high-DMI strip. This self-recovery mechanism is illustrated in Figs.~\ref{fig:selfrecovery}(a)-\ref{fig:selfrecovery}(c). Note that the geometrically designed ferromagnetic strips for domain-wall based waveguides do not have this self-healing property: Once the domain wall is deformed in such a way that it crosses the boundary of the waveguide, it will not relax back to the state with a single domain wall at the center of the strip [Figs.~\ref{fig:selfrecovery}(d)-\ref{fig:selfrecovery}(f)].

\subsection{\label{sec:racetrack}Racing skyrmions on high-DMI tracks}

A spin-polarized current can move skyrmions along a racetrack. Due to drifting and, more importantly, the skyrmion Hall effect, it is possible for the skyrmion to leave the track. This process, however, has to overcome an energy barrier related to the repulsive force between the skyrmion and the edge of the racetrack. For strong spin currents, the skyrmion Hall effect can become sufficiently large to expel the skyrmion out of the racetrack.

\begin{figure}[t!]
    \centering
    \includegraphics{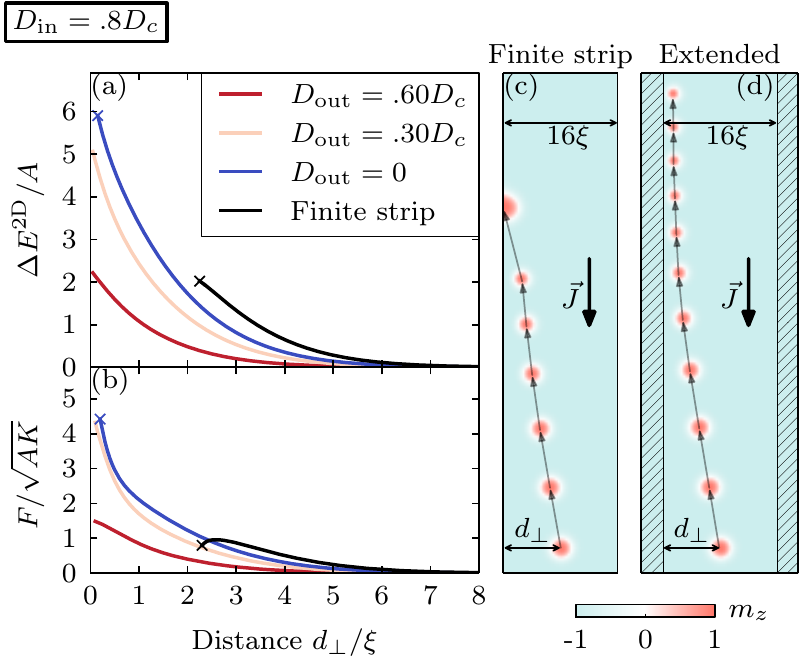} 
    \caption{Repulsion energy (a) of a skyrmion at a distance~$d_{\bot}$ from the edge of the racetrack and the corresponding force (b), for two design methods of the track (finite chiral strip versus the high-DMI strip within the extended film with lower DMI). A comparative snapshot of racing skyrmions for the two designs (with same DMI $D_{\mathrm{in}}=0.80D_c$, width of the track $w=16\xi$, and spin-current density $J$) is shown in panels (c) and (d).}
    \label{fig:repulsion}
\end{figure}

Racetracks are usually designed by shape engineering of the ferromagnetic film. We propose an alternative method, in which the racetrack is created by a high-DMI strip in an extended ferromagnetic film, with weak or no DMI outside the strip. We compare the repulsive force between skyrmions and the track's edges for the two design methods, by calculating the energy of a skyrmion as a function of its distance~$d_{\bot}$ to the edge of the racetrack. This is done by relaxing the magnetization while keeping the magnetic moment at the center of the skyrmion fixed at a certain distance~$d_{\bot}$ from the edge. The obtained energies and repulsive forces are shown in Figs.~\ref{fig:repulsion}(a) and \ref{fig:repulsion}(b). When we subsequently relaxed the magnetization without fixing any spins, we see that the skyrmions move back to the center of the strip and they do not collapse, as long as there is a repulsive force with the edge. This proves that our fixed-spin method, used to determine the repulsive force, is justified.

A skyrmion placed close to the boundary of a finite chiral ferromagnet ($d_{\bot}<2.4\xi$ for $D=0.8D_c$) will exit through the boundary, as shown by the shorter energy/force curves in Figs.~\ref{fig:repulsion}(a) and \ref{fig:repulsion}(b) for the shape-engineered racetrack. In the case of our racetrack, designed by heterogeneous DMI, the skyrmion can also escape from the DMI track by moving into the low DMI regions. This, however, requires overcoming a much larger energy barrier, especially if $D_{\mathrm{out}}$ is much smaller than $D_{\mathrm{in}}$. Consequently, the high-DMI track can sustain larger spin currents, without losing the skyrmion, than the ordinary shape-engineered DMI track. Figures~\ref{fig:repulsion}(c) and \ref{fig:repulsion}(d) shows the path of a skyrmion in the two types of racetracks for the same spin current and inner DMI strength. For this current the skyrmion escapes from the shape-engineered track, whereas it keeps racing on the high-DMI track. This significantly improves the performance of the racetrack memory device and dramatically reduces its volatility to skyrmion collapse.

\section{\label{sec:conclusion}Conclusion}

To summarize, we demonstrated in this paper the new manner to manipulate chiral spin structures in ferromagnetic films, such as cycloids and skyrmions, by engineering spatially the Dzyaloshinskii-Moriya interaction (DMI). Besides the useful analytic considerations, we showed that domain walls and skyrmions can be very effectively confined inside the prepatterned regions with higher DMI and how the properties of the ground state depend on the width of the confinement and interfaces between the regions with different DMI in a heterochiral film. We propose to utilize these findings in the advanced design of devices based on spatial DMI engineering, such as curved spin waveguides, and devices requiring precise selectivity of the shape and size of magnetic domains. We also demonstrate a much improved functionality of a skyrmion racetrack memory for a track defined by a high-DMI strip within an extended film with lower (or no) DMI, due to much increased repulsive force between a skyrmion and the border where DMI changes compared to the force keeping the skyrmion within the conventional finite tracks. Since the interfacially-induced DMI in a ferromagnetic film is possible to spatially engineer in experiment by, e.g., patterning the adjacent heavy-metal layer, we expect that our findings are only the first of emergent phenomena to be revealed in heterochiral, ferromagnetic films in the coming years.

\appendix*
\section{}

\begin{figure}[b]
    \centering
    \includegraphics{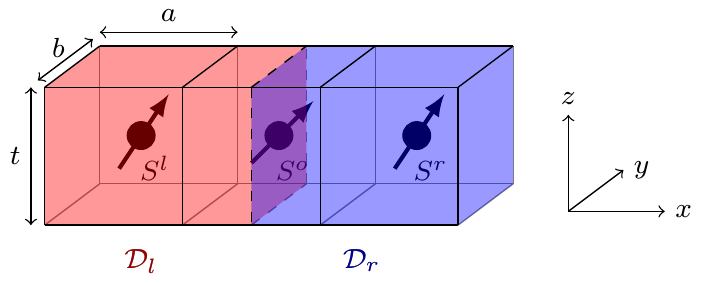}
    \caption{Cartoon of atomistic spin model with a magnetic moment~$S^o$ at an interface with on the left side DMI strength $\mathcal{D}_l$ and on the right side DMI strength~$\mathcal{D}_r$.}
    \label{fig:cells}
\end{figure}

In this Appendix, we demonstrate how one can translate the DMI energy expression of a classical atomistic Heisenberg model $\mathcal{D}_{ij}\cdot(S_i \times S_j)$ to the DMI energy density in a micromagnetic model, in the case of a spatially varying DMI strength. Let us consider a magnetic moment $S^o$ at an interface between DMI strength $\mathcal{D}_l$ on the left side and $\mathcal{D}_r$ on the right side, as depicted in Fig.~\ref{fig:cells}. The DMI energy density $\varepsilon_{\mathrm{dmi}}^o$ at this position depends on the magnetic moment on the left $S^l$ and the one on the right $S^r$, both at an interatomic distance~$a$ (assuming constant magnetization along the $y$ direction). The DMI energy density in the cell at the interface, with volume~$V_{\mathrm{cell}} = a b t$, then becomes
\begin{equation}
    \varepsilon_{\mathrm{dmi}}^o = \frac{1}{2 a b t} \left[ \mathcal{D}_l \left( S^o \times S^l \right)_y - \mathcal{D}_r \left( S^o \times S^r \right)_y \right]. \\
\end{equation}
One easily finds that this can be rewritten as
\begin{multline}
    \varepsilon_{\mathrm{dmi}}^o = \frac{1}{2 b t} \left[ S^o_x ( \mathcal{D}_l \delta^l S^o_z + \mathcal{D}_r \delta^r S^o_z ) \right. \\
    \left. - S^o_z ( \mathcal{D}_l \delta^l S^o_x + \mathcal{D}_r \delta^r S^o_x ) \right], \label{eq:atomic}
\end{multline}
with
\[
    \delta^l S^o = \frac{S^o-S^l}{a} \quad \mathrm{and} \quad \delta^r S^o = \frac{S^r-S^o}{a}
\]
which can be considered as the first-order left and right finite difference of the magnetization field $m$ at the interface. Note that expression~(\ref{eq:atomic}) is equivalent with expression~(\ref{eq:dmidens}) of the DMI energy density in the micromagnetic model where we allow for different left and right derivative. Note that if $\mathcal{D}_l=\mathcal{D}_r$, then the sum of two derivatives yields the second-order central difference. We therefore conclude that the energy densities from the two models are equivalent up to first order at the interface, and up to second order elsewhere.

\begin{acknowledgements}
This work was supported by the Fonds Wetenschappelijk Onderzoek (FWO-Vlaanderen) through Project No. G098917N.
\end{acknowledgements}

\bibliography{main.auto,coey}

\begin{thebibliography}{40}%
\makeatletter
\providecommand \@ifxundefined [1]{%
 \@ifx{#1\undefined}
}%
\providecommand \@ifnum [1]{%
 \ifnum #1\expandafter \@firstoftwo
 \else \expandafter \@secondoftwo
 \fi
}%
\providecommand \@ifx [1]{%
 \ifx #1\expandafter \@firstoftwo
 \else \expandafter \@secondoftwo
 \fi
}%
\providecommand \natexlab [1]{#1}%
\providecommand \enquote  [1]{``#1''}%
\providecommand \bibnamefont  [1]{#1}%
\providecommand \bibfnamefont [1]{#1}%
\providecommand \citenamefont [1]{#1}%
\providecommand \href@noop [0]{\@secondoftwo}%
\providecommand \href [0]{\begingroup \@sanitize@url \@href}%
\providecommand \@href[1]{\@@startlink{#1}\@@href}%
\providecommand \@@href[1]{\endgroup#1\@@endlink}%
\providecommand \@sanitize@url [0]{\catcode `\\12\catcode `\$12\catcode
  `\&12\catcode `\#12\catcode `\^12\catcode `\_12\catcode `\%12\relax}%
\providecommand \@@startlink[1]{}%
\providecommand \@@endlink[0]{}%
\providecommand \url  [0]{\begingroup\@sanitize@url \@url }%
\providecommand \@url [1]{\endgroup\@href {#1}{\urlprefix }}%
\providecommand \urlprefix  [0]{URL }%
\providecommand \Eprint [0]{\href }%
\providecommand \doibase [0]{http://dx.doi.org/}%
\providecommand \selectlanguage [0]{\@gobble}%
\providecommand \bibinfo  [0]{\@secondoftwo}%
\providecommand \bibfield  [0]{\@secondoftwo}%
\providecommand \translation [1]{[#1]}%
\providecommand \BibitemOpen [0]{}%
\providecommand \bibitemStop [0]{}%
\providecommand \bibitemNoStop [0]{.\EOS\space}%
\providecommand \EOS [0]{\spacefactor3000\relax}%
\providecommand \BibitemShut  [1]{\csname bibitem#1\endcsname}%
\let\auto@bib@innerbib\@empty
\bibitem [{\citenamefont {Dzyaloshinsky}(1958)}]{Dzyaloshinsky1958}%
  \BibitemOpen
  \bibfield  {author} {\bibinfo {author} {\bibfnamefont {I.~E.}\ \bibnamefont
  {Dzyaloshinsky}},\ }\href {\doibase 10.1016/0022-3697(58)90076-3} {\bibfield
  {journal} {\bibinfo  {journal} {J. Phys. Chem. Solids}\ }\textbf {\bibinfo
  {volume} {4}},\ \bibinfo {pages} {241} (\bibinfo {year} {1958})}\BibitemShut
  {NoStop}%
\bibitem [{\citenamefont {Moriya}(1960)}]{Moriya1960}%
  \BibitemOpen
  \bibfield  {author} {\bibinfo {author} {\bibfnamefont {T.~T.}\ \bibnamefont
  {Moriya}},\ }\href {\doibase 10.1103/PhysRev.120.91} {\bibfield  {journal}
  {\bibinfo  {journal} {Physical Review}\ }\textbf {\bibinfo {volume} {120}},\
  \bibinfo {pages} {91} (\bibinfo {year} {1960})}\BibitemShut {NoStop}%
\bibitem [{\citenamefont {Dzyaloshinskii}(1964)}]{Dzyaloshinsky1964}%
  \BibitemOpen
  \bibfield  {author} {\bibinfo {author} {\bibfnamefont {I.~E.}\ \bibnamefont
  {Dzyaloshinskii}},\ }\href
  {http://www.jetp.ac.ru/cgi-bin/e/index/e/19/4/p960?a=list} {\bibfield
  {journal} {\bibinfo  {journal} {Sov. Phys. JETP}\ }\textbf {\bibinfo {volume}
  {19}},\ \bibinfo {pages} {960} (\bibinfo {year} {1964})}\BibitemShut
  {NoStop}%
\bibitem [{\citenamefont {Cr{\'{e}}pieux}\ and\ \citenamefont
  {Lacroix}(1998)}]{Crepieux1998}%
  \BibitemOpen
  \bibfield  {author} {\bibinfo {author} {\bibfnamefont {A.}~\bibnamefont
  {Cr{\'{e}}pieux}}\ and\ \bibinfo {author} {\bibfnamefont {C.}~\bibnamefont
  {Lacroix}},\ }\href {\doibase 10.1016/S0304-8853(97)01044-5} {\bibfield
  {journal} {\bibinfo  {journal} {Journal of Magnetism and Magnetic Materials}\
  }\textbf {\bibinfo {volume} {182}},\ \bibinfo {pages} {341} (\bibinfo {year}
  {1998})}\BibitemShut {NoStop}%
\bibitem [{\citenamefont {Bogdanov}\ and\ \citenamefont
  {Hubert}(1994)}]{Bogdanov1994}%
  \BibitemOpen
  \bibfield  {author} {\bibinfo {author} {\bibfnamefont {A.}~\bibnamefont
  {Bogdanov}}\ and\ \bibinfo {author} {\bibfnamefont {A.}~\bibnamefont
  {Hubert}},\ }\href {\doibase 10.1016/0304-8853(94)90046-9} {\bibfield
  {journal} {\bibinfo  {journal} {Journal of Magnetism and Magnetic Materials}\
  }\textbf {\bibinfo {volume} {138}},\ \bibinfo {pages} {255} (\bibinfo {year}
  {1994})}\BibitemShut {NoStop}%
\bibitem [{\citenamefont {Bogdanov}\ and\ \citenamefont
  {R{\"{o}}{\ss}ler}(2001)}]{Bogdanov2001}%
  \BibitemOpen
  \bibfield  {author} {\bibinfo {author} {\bibfnamefont {A.~N.}\ \bibnamefont
  {Bogdanov}}\ and\ \bibinfo {author} {\bibfnamefont {U.~K.}\ \bibnamefont
  {R{\"{o}}{\ss}ler}},\ }\href {\doibase 10.1103/PhysRevLett.87.037203}
  {\bibfield  {journal} {\bibinfo  {journal} {Physical Review Letters}\
  }\textbf {\bibinfo {volume} {87}},\ \bibinfo {pages} {037203} (\bibinfo
  {year} {2001})}\BibitemShut {NoStop}%
\bibitem [{\citenamefont {Ezawa}(2011)}]{Ezawa2011}%
  \BibitemOpen
  \bibfield  {author} {\bibinfo {author} {\bibfnamefont {M.}~\bibnamefont
  {Ezawa}},\ }\href {\doibase 10.1103/PhysRevB.83.100408} {\bibfield  {journal}
  {\bibinfo  {journal} {Physical Review B}\ }\textbf {\bibinfo {volume} {83}},\
  \bibinfo {pages} {100408} (\bibinfo {year} {2011})}\BibitemShut {NoStop}%
\bibitem [{\citenamefont {Kiselev}\ \emph {et~al.}(2011)\citenamefont
  {Kiselev}, \citenamefont {Bogdanov}, \citenamefont {Sch{\"{a}}fer},\ and\
  \citenamefont {R{\"{o}}{\ss}ler}}]{Kiselev2011}%
  \BibitemOpen
  \bibfield  {author} {\bibinfo {author} {\bibfnamefont {N.~S.}\ \bibnamefont
  {Kiselev}}, \bibinfo {author} {\bibfnamefont {A.~N.}\ \bibnamefont
  {Bogdanov}}, \bibinfo {author} {\bibfnamefont {R.}~\bibnamefont
  {Sch{\"{a}}fer}}, \ and\ \bibinfo {author} {\bibfnamefont {U.~K.}\
  \bibnamefont {R{\"{o}}{\ss}ler}},\ }\href {\doibase
  10.1088/0022-3727/44/39/392001} {\bibfield  {journal} {\bibinfo  {journal}
  {Journal of Physics D: Applied Physics}\ }\textbf {\bibinfo {volume} {44}},\
  \bibinfo {pages} {392001} (\bibinfo {year} {2011})}\BibitemShut {NoStop}%
\bibitem [{\citenamefont {Rohart}\ and\ \citenamefont
  {Thiaville}(2013)}]{Rohart2013}%
  \BibitemOpen
  \bibfield  {author} {\bibinfo {author} {\bibfnamefont {S.}~\bibnamefont
  {Rohart}}\ and\ \bibinfo {author} {\bibfnamefont {A.}~\bibnamefont
  {Thiaville}},\ }\href {\doibase 10.1103/PhysRevB.88.184422} {\bibfield
  {journal} {\bibinfo  {journal} {Physical Review B}\ }\textbf {\bibinfo
  {volume} {88}},\ \bibinfo {pages} {184422} (\bibinfo {year}
  {2013})}\BibitemShut {NoStop}%
\bibitem [{\citenamefont {Chui}\ \emph {et~al.}(2015)\citenamefont {Chui},
  \citenamefont {Ma},\ and\ \citenamefont {Zhou}}]{Chui2015}%
  \BibitemOpen
  \bibfield  {author} {\bibinfo {author} {\bibfnamefont {C.~P.}\ \bibnamefont
  {Chui}}, \bibinfo {author} {\bibfnamefont {F.}~\bibnamefont {Ma}}, \ and\
  \bibinfo {author} {\bibfnamefont {Y.}~\bibnamefont {Zhou}},\ }\href {\doibase
  10.1063/1.4919320} {\bibfield  {journal} {\bibinfo  {journal} {AIP Advances}\
  }\textbf {\bibinfo {volume} {5}},\ \bibinfo {pages} {047141} (\bibinfo {year}
  {2015})}\BibitemShut {NoStop}%
\bibitem [{\citenamefont {Mulkers}\ \emph {et~al.}(2016)\citenamefont
  {Mulkers}, \citenamefont {Milo{\v{s}}evi{\'{c}}},\ and\ \citenamefont {{Van
  Waeyenberge}}}]{Mulkers2016}%
  \BibitemOpen
  \bibfield  {author} {\bibinfo {author} {\bibfnamefont {J.}~\bibnamefont
  {Mulkers}}, \bibinfo {author} {\bibfnamefont {M.~V.}\ \bibnamefont
  {Milo{\v{s}}evi{\'{c}}}}, \ and\ \bibinfo {author} {\bibfnamefont
  {B.}~\bibnamefont {{Van Waeyenberge}}},\ }\href {\doibase
  10.1103/PhysRevB.93.214405} {\bibfield  {journal} {\bibinfo  {journal}
  {Physical Review B}\ }\textbf {\bibinfo {volume} {93}},\ \bibinfo {pages}
  {214405} (\bibinfo {year} {2016})}\BibitemShut {NoStop}%
\bibitem [{\citenamefont {Navau}\ \emph {et~al.}(2016)\citenamefont {Navau},
  \citenamefont {Del-Valle},\ and\ \citenamefont {Sanchez}}]{Navau2016}%
  \BibitemOpen
  \bibfield  {author} {\bibinfo {author} {\bibfnamefont {C.}~\bibnamefont
  {Navau}}, \bibinfo {author} {\bibfnamefont {N.}~\bibnamefont {Del-Valle}}, \
  and\ \bibinfo {author} {\bibfnamefont {A.}~\bibnamefont {Sanchez}},\ }\href
  {\doibase 10.1103/PhysRevB.94.184104} {\bibfield  {journal} {\bibinfo
  {journal} {Physical Review B}\ }\textbf {\bibinfo {volume} {94}},\ \bibinfo
  {pages} {184104} (\bibinfo {year} {2016})}\BibitemShut {NoStop}%
\bibitem [{\citenamefont {Fert}\ \emph {et~al.}(2013)\citenamefont {Fert},
  \citenamefont {Cros},\ and\ \citenamefont {Sampaio}}]{Fert2013}%
  \BibitemOpen
  \bibfield  {author} {\bibinfo {author} {\bibfnamefont {A.}~\bibnamefont
  {Fert}}, \bibinfo {author} {\bibfnamefont {V.}~\bibnamefont {Cros}}, \ and\
  \bibinfo {author} {\bibfnamefont {J.}~\bibnamefont {Sampaio}},\ }\href
  {\doibase 10.1038/nnano.2013.29} {\bibfield  {journal} {\bibinfo  {journal}
  {Nature Nanotechnology}\ }\textbf {\bibinfo {volume} {8}},\ \bibinfo {pages}
  {152} (\bibinfo {year} {2013})}\BibitemShut {NoStop}%
\bibitem [{\citenamefont {Tomasello}\ \emph {et~al.}(2014)\citenamefont
  {Tomasello}, \citenamefont {Martinez}, \citenamefont {Zivieri}, \citenamefont
  {Torres}, \citenamefont {Carpentieri},\ and\ \citenamefont
  {Finocchio}}]{tomasello2014}%
  \BibitemOpen
  \bibfield  {author} {\bibinfo {author} {\bibfnamefont {R.}~\bibnamefont
  {Tomasello}}, \bibinfo {author} {\bibfnamefont {E.}~\bibnamefont {Martinez}},
  \bibinfo {author} {\bibfnamefont {R.}~\bibnamefont {Zivieri}}, \bibinfo
  {author} {\bibfnamefont {L.}~\bibnamefont {Torres}}, \bibinfo {author}
  {\bibfnamefont {M.}~\bibnamefont {Carpentieri}}, \ and\ \bibinfo {author}
  {\bibfnamefont {G.}~\bibnamefont {Finocchio}},\ }\href {\doibase
  10.1038/srep06784} {\bibfield  {journal} {\bibinfo  {journal} {Scientific
  Reports}\ }\textbf {\bibinfo {volume} {4}},\ \bibinfo {pages} {6784}
  (\bibinfo {year} {2014})}\BibitemShut {NoStop}%
\bibitem [{\citenamefont {Zhang}\ \emph
  {et~al.}(2015{\natexlab{a}})\citenamefont {Zhang}, \citenamefont {Zhao},
  \citenamefont {Fangohr}, \citenamefont {Liu}, \citenamefont {Xia},
  \citenamefont {Xia},\ and\ \citenamefont {Morvan}}]{Zhang2015}%
  \BibitemOpen
  \bibfield  {author} {\bibinfo {author} {\bibfnamefont {X.}~\bibnamefont
  {Zhang}}, \bibinfo {author} {\bibfnamefont {G.~P.}\ \bibnamefont {Zhao}},
  \bibinfo {author} {\bibfnamefont {H.}~\bibnamefont {Fangohr}}, \bibinfo
  {author} {\bibfnamefont {J.~P.}\ \bibnamefont {Liu}}, \bibinfo {author}
  {\bibfnamefont {W.~X.}\ \bibnamefont {Xia}}, \bibinfo {author} {\bibfnamefont
  {J.}~\bibnamefont {Xia}}, \ and\ \bibinfo {author} {\bibfnamefont {F.~J.}\
  \bibnamefont {Morvan}},\ }\href {\doibase 10.1038/srep07643} {\bibfield
  {journal} {\bibinfo  {journal} {Scientific Reports}\ }\textbf {\bibinfo
  {volume} {5}},\ \bibinfo {pages} {7643} (\bibinfo {year}
  {2015}{\natexlab{a}})}\BibitemShut {NoStop}%
\bibitem [{\citenamefont {Iwasaki}\ \emph {et~al.}(2014)\citenamefont
  {Iwasaki}, \citenamefont {Koshibae},\ and\ \citenamefont
  {Nagaosa}}]{Iwasaki2014}%
  \BibitemOpen
  \bibfield  {author} {\bibinfo {author} {\bibfnamefont {J.}~\bibnamefont
  {Iwasaki}}, \bibinfo {author} {\bibfnamefont {W.}~\bibnamefont {Koshibae}}, \
  and\ \bibinfo {author} {\bibfnamefont {N.}~\bibnamefont {Nagaosa}},\ }\href
  {\doibase 10.1021/nl501379k} {\bibfield  {journal} {\bibinfo  {journal} {Nano
  Letters}\ }\textbf {\bibinfo {volume} {14}},\ \bibinfo {pages} {4432}
  (\bibinfo {year} {2014})}\BibitemShut {NoStop}%
\bibitem [{\citenamefont {Zhang}\ \emph
  {et~al.}(2015{\natexlab{b}})\citenamefont {Zhang}, \citenamefont {Zhou},
  \citenamefont {Ezawa}, \citenamefont {Zhao},\ and\ \citenamefont
  {Zhao}}]{Zhang2015b}%
  \BibitemOpen
  \bibfield  {author} {\bibinfo {author} {\bibfnamefont {X.}~\bibnamefont
  {Zhang}}, \bibinfo {author} {\bibfnamefont {Y.}~\bibnamefont {Zhou}},
  \bibinfo {author} {\bibfnamefont {M.}~\bibnamefont {Ezawa}}, \bibinfo
  {author} {\bibfnamefont {G.~P.}\ \bibnamefont {Zhao}}, \ and\ \bibinfo
  {author} {\bibfnamefont {W.}~\bibnamefont {Zhao}},\ }\href {\doibase
  10.1038/srep11369} {\bibfield  {journal} {\bibinfo  {journal} {Scientific
  Reports}\ }\textbf {\bibinfo {volume} {5}},\ \bibinfo {pages} {11369}
  (\bibinfo {year} {2015}{\natexlab{b}})}\BibitemShut {NoStop}%
\bibitem [{\citenamefont {Kang}\ \emph {et~al.}(2016)\citenamefont {Kang},
  \citenamefont {Huang}, \citenamefont {Zheng}, \citenamefont {Lv},
  \citenamefont {Lei}, \citenamefont {Zhang}, \citenamefont {Zhang},
  \citenamefont {Zhou},\ and\ \citenamefont {Zhao}}]{Kang2016}%
  \BibitemOpen
  \bibfield  {author} {\bibinfo {author} {\bibfnamefont {W.}~\bibnamefont
  {Kang}}, \bibinfo {author} {\bibfnamefont {Y.}~\bibnamefont {Huang}},
  \bibinfo {author} {\bibfnamefont {C.}~\bibnamefont {Zheng}}, \bibinfo
  {author} {\bibfnamefont {W.}~\bibnamefont {Lv}}, \bibinfo {author}
  {\bibfnamefont {N.}~\bibnamefont {Lei}}, \bibinfo {author} {\bibfnamefont
  {Y.}~\bibnamefont {Zhang}}, \bibinfo {author} {\bibfnamefont
  {X.}~\bibnamefont {Zhang}}, \bibinfo {author} {\bibfnamefont
  {Y.}~\bibnamefont {Zhou}}, \ and\ \bibinfo {author} {\bibfnamefont
  {W.}~\bibnamefont {Zhao}},\ }\href {\doibase 10.1038/srep23164} {\bibfield
  {journal} {\bibinfo  {journal} {Scientific Reports}\ }\textbf {\bibinfo
  {volume} {6}},\ \bibinfo {pages} {23164} (\bibinfo {year}
  {2016})}\BibitemShut {NoStop}%
\bibitem [{\citenamefont {Fook}\ \emph {et~al.}(2016)\citenamefont {Fook},
  \citenamefont {Gan},\ and\ \citenamefont {Lew}}]{Fook2016}%
  \BibitemOpen
  \bibfield  {author} {\bibinfo {author} {\bibfnamefont {H.~T.}\ \bibnamefont
  {Fook}}, \bibinfo {author} {\bibfnamefont {W.~L.}\ \bibnamefont {Gan}}, \
  and\ \bibinfo {author} {\bibfnamefont {W.~S.}\ \bibnamefont {Lew}},\ }\href
  {\doibase 10.1038/srep21099} {\bibfield  {journal} {\bibinfo  {journal}
  {Scientific Reports}\ }\textbf {\bibinfo {volume} {6}},\ \bibinfo {pages}
  {21099} (\bibinfo {year} {2016})}\BibitemShut {NoStop}%
\bibitem [{\citenamefont {Hrabec}\ \emph {et~al.}(2014)\citenamefont {Hrabec},
  \citenamefont {Porter}, \citenamefont {Wells}, \citenamefont {Benitez},
  \citenamefont {Burnell}, \citenamefont {McVitie}, \citenamefont {McGrouther},
  \citenamefont {Moore},\ and\ \citenamefont {Marrows}}]{Hrabec2014}%
  \BibitemOpen
  \bibfield  {author} {\bibinfo {author} {\bibfnamefont {A.}~\bibnamefont
  {Hrabec}}, \bibinfo {author} {\bibfnamefont {N.~A.}\ \bibnamefont {Porter}},
  \bibinfo {author} {\bibfnamefont {A.}~\bibnamefont {Wells}}, \bibinfo
  {author} {\bibfnamefont {M.~J.}\ \bibnamefont {Benitez}}, \bibinfo {author}
  {\bibfnamefont {G.}~\bibnamefont {Burnell}}, \bibinfo {author} {\bibfnamefont
  {S.}~\bibnamefont {McVitie}}, \bibinfo {author} {\bibfnamefont
  {D.}~\bibnamefont {McGrouther}}, \bibinfo {author} {\bibfnamefont {T.~A.}\
  \bibnamefont {Moore}}, \ and\ \bibinfo {author} {\bibfnamefont {C.~H.}\
  \bibnamefont {Marrows}},\ }\href {\doibase 10.1103/PhysRevB.90.020402}
  {\bibfield  {journal} {\bibinfo  {journal} {Physical Review B - Condensed
  Matter and Materials Physics}\ }\textbf {\bibinfo {volume} {90}},\ \bibinfo
  {pages} {020402} (\bibinfo {year} {2014})}\BibitemShut {NoStop}%
\bibitem [{\citenamefont {Kim}\ \emph {et~al.}(2015)\citenamefont {Kim},
  \citenamefont {Han}, \citenamefont {Jung}, \citenamefont {Cho}, \citenamefont
  {Kim}, \citenamefont {Swagten},\ and\ \citenamefont {You}}]{Kim2015}%
  \BibitemOpen
  \bibfield  {author} {\bibinfo {author} {\bibfnamefont {N.-H.}\ \bibnamefont
  {Kim}}, \bibinfo {author} {\bibfnamefont {D.-S.}\ \bibnamefont {Han}},
  \bibinfo {author} {\bibfnamefont {J.}~\bibnamefont {Jung}}, \bibinfo {author}
  {\bibfnamefont {J.}~\bibnamefont {Cho}}, \bibinfo {author} {\bibfnamefont
  {J.-S.}\ \bibnamefont {Kim}}, \bibinfo {author} {\bibfnamefont {H.~J.~M.}\
  \bibnamefont {Swagten}}, \ and\ \bibinfo {author} {\bibfnamefont {C.~Y.}\
  \bibnamefont {You}},\ }\href {\doibase 10.1063/1.4932550} {\bibfield
  {journal} {\bibinfo  {journal} {Applied Physics Letters}\ }\textbf {\bibinfo
  {volume} {107}},\ \bibinfo {pages} {142408} (\bibinfo {year}
  {2015})}\BibitemShut {NoStop}%
\bibitem [{\citenamefont {Cho}\ \emph {et~al.}(2015)\citenamefont {Cho},
  \citenamefont {Kim}, \citenamefont {Lee}, \citenamefont {Kim}, \citenamefont
  {Lavrijsen}, \citenamefont {Solignac}, \citenamefont {Yin}, \citenamefont
  {Han}, \citenamefont {van Hoof}, \citenamefont {Swagten}, \citenamefont
  {Koopmans},\ and\ \citenamefont {You}}]{Cho2015}%
  \BibitemOpen
  \bibfield  {author} {\bibinfo {author} {\bibfnamefont {J.}~\bibnamefont
  {Cho}}, \bibinfo {author} {\bibfnamefont {N.-H.}\ \bibnamefont {Kim}},
  \bibinfo {author} {\bibfnamefont {S.}~\bibnamefont {Lee}}, \bibinfo {author}
  {\bibfnamefont {J.-S.}\ \bibnamefont {Kim}}, \bibinfo {author} {\bibfnamefont
  {R.}~\bibnamefont {Lavrijsen}}, \bibinfo {author} {\bibfnamefont
  {A.}~\bibnamefont {Solignac}}, \bibinfo {author} {\bibfnamefont
  {Y.}~\bibnamefont {Yin}}, \bibinfo {author} {\bibfnamefont {D.-S.}\
  \bibnamefont {Han}}, \bibinfo {author} {\bibfnamefont {N.~J.~J.}\
  \bibnamefont {van Hoof}}, \bibinfo {author} {\bibfnamefont {H.~J.~M.}\
  \bibnamefont {Swagten}}, \bibinfo {author} {\bibfnamefont {B.}~\bibnamefont
  {Koopmans}}, \ and\ \bibinfo {author} {\bibfnamefont {C.-Y.}\ \bibnamefont
  {You}},\ }\href {\doibase 10.1038/ncomms8635} {\bibfield  {journal} {\bibinfo
   {journal} {Nature Communications}\ }\textbf {\bibinfo {volume} {6}},\
  \bibinfo {pages} {7635} (\bibinfo {year} {2015})}\BibitemShut {NoStop}%
\bibitem [{\citenamefont {Belmeguenai}\ \emph {et~al.}(2016)\citenamefont
  {Belmeguenai}, \citenamefont {Gabor}, \citenamefont {Roussign{\'{e}}},
  \citenamefont {Stashkevich}, \citenamefont {Ch{\'{e}}rif}, \citenamefont
  {Zighem},\ and\ \citenamefont {Tiusan}}]{Belmeguenai2016}%
  \BibitemOpen
  \bibfield  {author} {\bibinfo {author} {\bibfnamefont {M.}~\bibnamefont
  {Belmeguenai}}, \bibinfo {author} {\bibfnamefont {M.~S.}\ \bibnamefont
  {Gabor}}, \bibinfo {author} {\bibfnamefont {Y.}~\bibnamefont
  {Roussign{\'{e}}}}, \bibinfo {author} {\bibfnamefont {A.}~\bibnamefont
  {Stashkevich}}, \bibinfo {author} {\bibfnamefont {S.~M.}\ \bibnamefont
  {Ch{\'{e}}rif}}, \bibinfo {author} {\bibfnamefont {F.}~\bibnamefont
  {Zighem}}, \ and\ \bibinfo {author} {\bibfnamefont {C.}~\bibnamefont
  {Tiusan}},\ }\href {\doibase 10.1103/PhysRevB.93.174407} {\bibfield
  {journal} {\bibinfo  {journal} {Physical Review B}\ }\textbf {\bibinfo
  {volume} {93}},\ \bibinfo {pages} {174407} (\bibinfo {year}
  {2016})}\BibitemShut {NoStop}%
\bibitem [{\citenamefont {Yang}\ \emph {et~al.}(2016)\citenamefont {Yang},
  \citenamefont {Boulle}, \citenamefont {Cros}, \citenamefont {Fert},\ and\
  \citenamefont {Chshiev}}]{Yang}%
  \BibitemOpen
  \bibfield  {author} {\bibinfo {author} {\bibfnamefont {H.}~\bibnamefont
  {Yang}}, \bibinfo {author} {\bibfnamefont {O.}~\bibnamefont {Boulle}},
  \bibinfo {author} {\bibfnamefont {V.}~\bibnamefont {Cros}}, \bibinfo {author}
  {\bibfnamefont {A.}~\bibnamefont {Fert}}, \ and\ \bibinfo {author}
  {\bibfnamefont {M.}~\bibnamefont {Chshiev}},\ }\href
  {http://arxiv.org/abs/1603.01847} {\bibfield  {journal} {\bibinfo  {journal}
  {Arxiv Preprint}\ ,\ \bibinfo {pages} {1603.01847}} (\bibinfo {year}
  {2016})},\ \Eprint {http://arxiv.org/abs/1603.01847} {arXiv:1603.01847}
  \BibitemShut {NoStop}%
\bibitem [{\citenamefont {Tacchi}\ \emph {et~al.}(2016)\citenamefont {Tacchi},
  \citenamefont {Troncoso}, \citenamefont {Ahlberg}, \citenamefont {Gubbiotti},
  \citenamefont {Madami}, \citenamefont {{\AA}kerman},\ and\ \citenamefont
  {Landeros}}]{Tacchi2016}%
  \BibitemOpen
  \bibfield  {author} {\bibinfo {author} {\bibfnamefont {S.}~\bibnamefont
  {Tacchi}}, \bibinfo {author} {\bibfnamefont {R.~E.}\ \bibnamefont
  {Troncoso}}, \bibinfo {author} {\bibfnamefont {M.}~\bibnamefont {Ahlberg}},
  \bibinfo {author} {\bibfnamefont {G.}~\bibnamefont {Gubbiotti}}, \bibinfo
  {author} {\bibfnamefont {M.}~\bibnamefont {Madami}}, \bibinfo {author}
  {\bibfnamefont {J.}~\bibnamefont {{\AA}kerman}}, \ and\ \bibinfo {author}
  {\bibfnamefont {P.}~\bibnamefont {Landeros}},\ }\href
  {http://arxiv.org/abs/1604.02626} {\bibfield  {journal} {\bibinfo  {journal}
  {Arxiv Preprint}\ ,\ \bibinfo {pages} {1604.02626}} (\bibinfo {year}
  {2016})},\ \Eprint {http://arxiv.org/abs/1604.02626} {arXiv:1604.02626}
  \BibitemShut {NoStop}%
\bibitem [{\citenamefont {Ma}\ \emph {et~al.}(2016)\citenamefont {Ma},
  \citenamefont {Yu}, \citenamefont {Li}, \citenamefont {Wang}, \citenamefont
  {Wu}, \citenamefont {Olsson}, \citenamefont {Chu}, \citenamefont {An},
  \citenamefont {Xiao}, \citenamefont {Wang},\ and\ \citenamefont
  {Li}}]{Ma2016}%
  \BibitemOpen
  \bibfield  {author} {\bibinfo {author} {\bibfnamefont {X.}~\bibnamefont
  {Ma}}, \bibinfo {author} {\bibfnamefont {G.}~\bibnamefont {Yu}}, \bibinfo
  {author} {\bibfnamefont {X.}~\bibnamefont {Li}}, \bibinfo {author}
  {\bibfnamefont {T.}~\bibnamefont {Wang}}, \bibinfo {author} {\bibfnamefont
  {D.}~\bibnamefont {Wu}}, \bibinfo {author} {\bibfnamefont {K.~S.}\
  \bibnamefont {Olsson}}, \bibinfo {author} {\bibfnamefont {Z.}~\bibnamefont
  {Chu}}, \bibinfo {author} {\bibfnamefont {K.}~\bibnamefont {An}}, \bibinfo
  {author} {\bibfnamefont {J.~Q.}\ \bibnamefont {Xiao}}, \bibinfo {author}
  {\bibfnamefont {K.~L.}\ \bibnamefont {Wang}}, \ and\ \bibinfo {author}
  {\bibfnamefont {X.}~\bibnamefont {Li}},\ }\href {\doibase
  10.1103/PhysRevB.94.180408} {\bibfield  {journal} {\bibinfo  {journal}
  {Physical Review B - Condensed Matter and Materials Physics}\ }\textbf
  {\bibinfo {volume} {94}},\ \bibinfo {pages} {180408} (\bibinfo {year}
  {2016})}\BibitemShut {NoStop}%
\bibitem [{\citenamefont {Balk}\ \emph {et~al.}(2016)\citenamefont {Balk},
  \citenamefont {Kim}, \citenamefont {Pierce}, \citenamefont {Stiles},
  \citenamefont {Unguris},\ and\ \citenamefont {Stavis}}]{Balk2016}%
  \BibitemOpen
  \bibfield  {author} {\bibinfo {author} {\bibfnamefont {A.~L.}\ \bibnamefont
  {Balk}}, \bibinfo {author} {\bibfnamefont {K.-W.}\ \bibnamefont {Kim}},
  \bibinfo {author} {\bibfnamefont {D.~T.}\ \bibnamefont {Pierce}}, \bibinfo
  {author} {\bibfnamefont {M.~D.}\ \bibnamefont {Stiles}}, \bibinfo {author}
  {\bibfnamefont {J.}~\bibnamefont {Unguris}}, \ and\ \bibinfo {author}
  {\bibfnamefont {S.~M.}\ \bibnamefont {Stavis}},\ }\href
  {http://arxiv.org/abs/1609.09790} {\bibfield  {journal} {\bibinfo  {journal}
  {Arxiv Preprint}\ } (\bibinfo {year} {2016})},\ \Eprint
  {http://arxiv.org/abs/1609.09790} {arXiv:1609.09790} \BibitemShut {NoStop}%
\bibitem [{\citenamefont {Wells}\ \emph {et~al.}(2017)\citenamefont {Wells},
  \citenamefont {Shepley}, \citenamefont {Marrows},\ and\ \citenamefont
  {Moore}}]{Wells2017}%
  \BibitemOpen
  \bibfield  {author} {\bibinfo {author} {\bibfnamefont {A.~W.~J.}\
  \bibnamefont {Wells}}, \bibinfo {author} {\bibfnamefont {P.~M.}\ \bibnamefont
  {Shepley}}, \bibinfo {author} {\bibfnamefont {C.~H.}\ \bibnamefont
  {Marrows}}, \ and\ \bibinfo {author} {\bibfnamefont {T.~A.}\ \bibnamefont
  {Moore}},\ }\href {\doibase 10.1103/PhysRevB.95.054428} {\bibfield  {journal}
  {\bibinfo  {journal} {Physical Review B - Condensed Matter and Materials
  Physics}\ }\textbf {\bibinfo {volume} {95}},\ \bibinfo {pages} {054428}
  (\bibinfo {year} {2017})}\BibitemShut {NoStop}%
\bibitem [{\citenamefont {Xing}\ and\ \citenamefont {Zhou}(2016)}]{Xing2016}%
  \BibitemOpen
  \bibfield  {author} {\bibinfo {author} {\bibfnamefont {X.}~\bibnamefont
  {Xing}}\ and\ \bibinfo {author} {\bibfnamefont {Y.}~\bibnamefont {Zhou}},\
  }\href {\doibase 10.1038/am.2016.25} {\bibfield  {journal} {\bibinfo
  {journal} {NPG Asia Materials}\ }\textbf {\bibinfo {volume} {8}},\ \bibinfo
  {pages} {e246} (\bibinfo {year} {2016})}\BibitemShut {NoStop}%
\bibitem [{\citenamefont {Garcia-Sanchez}\ \emph {et~al.}(2014)\citenamefont
  {Garcia-Sanchez}, \citenamefont {Borys}, \citenamefont {Vansteenkiste},
  \citenamefont {Kim},\ and\ \citenamefont {Stamps}}]{Garcia-sanchez2014}%
  \BibitemOpen
  \bibfield  {author} {\bibinfo {author} {\bibfnamefont {F.}~\bibnamefont
  {Garcia-Sanchez}}, \bibinfo {author} {\bibfnamefont {P.}~\bibnamefont
  {Borys}}, \bibinfo {author} {\bibfnamefont {A.}~\bibnamefont
  {Vansteenkiste}}, \bibinfo {author} {\bibfnamefont {J.-V.}\ \bibnamefont
  {Kim}}, \ and\ \bibinfo {author} {\bibfnamefont {R.~L.}\ \bibnamefont
  {Stamps}},\ }\href {\doibase 10.1103/PhysRevB.89.224408} {\bibfield
  {journal} {\bibinfo  {journal} {Physical Review B}\ }\textbf {\bibinfo
  {volume} {89}},\ \bibinfo {pages} {224408} (\bibinfo {year}
  {2014})}\BibitemShut {NoStop}%
\bibitem [{\citenamefont {Garcia-Sanchez}\ \emph {et~al.}(2015)\citenamefont
  {Garcia-Sanchez}, \citenamefont {Borys}, \citenamefont {Soucaille},
  \citenamefont {Adam}, \citenamefont {Stamps},\ and\ \citenamefont
  {Kim}}]{Garcia-Sanchez2015}%
  \BibitemOpen
  \bibfield  {author} {\bibinfo {author} {\bibfnamefont {F.}~\bibnamefont
  {Garcia-Sanchez}}, \bibinfo {author} {\bibfnamefont {P.}~\bibnamefont
  {Borys}}, \bibinfo {author} {\bibfnamefont {R.}~\bibnamefont {Soucaille}},
  \bibinfo {author} {\bibfnamefont {J.-P.}\ \bibnamefont {Adam}}, \bibinfo
  {author} {\bibfnamefont {R.~L.}\ \bibnamefont {Stamps}}, \ and\ \bibinfo
  {author} {\bibfnamefont {J.-V.}\ \bibnamefont {Kim}},\ }\href {\doibase
  10.1103/PhysRevLett.114.247206} {\bibfield  {journal} {\bibinfo  {journal}
  {Physical Review Letters}\ }\textbf {\bibinfo {volume} {114}},\ \bibinfo
  {pages} {247206} (\bibinfo {year} {2015})}\BibitemShut {NoStop}%
\bibitem [{\citenamefont {Borys}\ \emph {et~al.}(2016)\citenamefont {Borys},
  \citenamefont {Garcia-Sanchez}, \citenamefont {Kim},\ and\ \citenamefont
  {Stamps}}]{Borys2016}%
  \BibitemOpen
  \bibfield  {author} {\bibinfo {author} {\bibfnamefont {P.}~\bibnamefont
  {Borys}}, \bibinfo {author} {\bibfnamefont {F.}~\bibnamefont
  {Garcia-Sanchez}}, \bibinfo {author} {\bibfnamefont {J.-V.}\ \bibnamefont
  {Kim}}, \ and\ \bibinfo {author} {\bibfnamefont {R.~L.}\ \bibnamefont
  {Stamps}},\ }\href {\doibase 10.1002/aelm.201500202} {\bibfield  {journal}
  {\bibinfo  {journal} {Advanced Electronic Materials}\ }\textbf {\bibinfo
  {volume} {2}},\ \bibinfo {pages} {1500202} (\bibinfo {year}
  {2016})}\BibitemShut {NoStop}%
\bibitem [{\citenamefont {Coey}(2010)}]{Coey}%
  \BibitemOpen
  \bibfield  {author} {\bibinfo {author} {\bibfnamefont {J.~M.~D.}\
  \bibnamefont {Coey}},\ }\href {http://dx.doi.org/10.1017/CBO9780511845000}
  {\emph {\bibinfo {title} {Magnetism and Magnetic Materials}}}\ (\bibinfo
  {publisher} {Cambridge University Press},\ \bibinfo {year} {2010})\ p.\
  \bibinfo {pages} {168}\BibitemShut {NoStop}%
\bibitem [{\citenamefont {Thiaville}\ \emph {et~al.}(2012)\citenamefont
  {Thiaville}, \citenamefont {Rohart}, \citenamefont {Ju{\'{e}}}, \citenamefont
  {Cros},\ and\ \citenamefont {Fert}}]{Thiaville2012}%
  \BibitemOpen
  \bibfield  {author} {\bibinfo {author} {\bibfnamefont {A.}~\bibnamefont
  {Thiaville}}, \bibinfo {author} {\bibfnamefont {S.}~\bibnamefont {Rohart}},
  \bibinfo {author} {\bibfnamefont {{\'{E}}.}~\bibnamefont {Ju{\'{e}}}},
  \bibinfo {author} {\bibfnamefont {V.}~\bibnamefont {Cros}}, \ and\ \bibinfo
  {author} {\bibfnamefont {A.}~\bibnamefont {Fert}},\ }\href {\doibase
  10.1209/0295-5075/100/57002} {\bibfield  {journal} {\bibinfo  {journal} {EPL
  (Europhysics Letters)}\ }\textbf {\bibinfo {volume} {100}},\ \bibinfo {pages}
  {57002} (\bibinfo {year} {2012})},\ \Eprint {http://arxiv.org/abs/1211.5970}
  {arXiv:1211.5970} \BibitemShut {NoStop}%
\bibitem [{\citenamefont {Vansteenkiste}\ \emph {et~al.}(2014)\citenamefont
  {Vansteenkiste}, \citenamefont {Leliaert}, \citenamefont {Dvornik},
  \citenamefont {Helsen}, \citenamefont {Garcia-Sanchez},\ and\ \citenamefont
  {{Van Waeyenberge}}}]{Vansteenkiste2014}%
  \BibitemOpen
  \bibfield  {author} {\bibinfo {author} {\bibfnamefont {A.}~\bibnamefont
  {Vansteenkiste}}, \bibinfo {author} {\bibfnamefont {J.}~\bibnamefont
  {Leliaert}}, \bibinfo {author} {\bibfnamefont {M.}~\bibnamefont {Dvornik}},
  \bibinfo {author} {\bibfnamefont {M.}~\bibnamefont {Helsen}}, \bibinfo
  {author} {\bibfnamefont {F.}~\bibnamefont {Garcia-Sanchez}}, \ and\ \bibinfo
  {author} {\bibfnamefont {B.}~\bibnamefont {{Van Waeyenberge}}},\ }\href
  {\doibase 10.1063/1.4899186} {\bibfield  {journal} {\bibinfo  {journal} {AIP
  Advances}\ }\textbf {\bibinfo {volume} {4}},\ \bibinfo {pages} {107133}
  (\bibinfo {year} {2014})}\BibitemShut {NoStop}%
\bibitem [{\citenamefont {Rohart}\ \emph {et~al.}(2016)\citenamefont {Rohart},
  \citenamefont {Miltat},\ and\ \citenamefont {Thiaville}}]{Rohart2016}%
  \BibitemOpen
  \bibfield  {author} {\bibinfo {author} {\bibfnamefont {S.}~\bibnamefont
  {Rohart}}, \bibinfo {author} {\bibfnamefont {J.}~\bibnamefont {Miltat}}, \
  and\ \bibinfo {author} {\bibfnamefont {A.}~\bibnamefont {Thiaville}},\ }\href
  {\doibase 10.1103/PhysRevB.93.214412} {\bibfield  {journal} {\bibinfo
  {journal} {Physical Review B}\ }\textbf {\bibinfo {volume} {93}},\ \bibinfo
  {pages} {214412} (\bibinfo {year} {2016})}\BibitemShut {NoStop}%
\bibitem [{\citenamefont {Lobanov}\ \emph {et~al.}(2016)\citenamefont
  {Lobanov}, \citenamefont {J{\'{o}}nsson},\ and\ \citenamefont
  {Uzdin}}]{Lobanov2016}%
  \BibitemOpen
  \bibfield  {author} {\bibinfo {author} {\bibfnamefont {I.~S.}\ \bibnamefont
  {Lobanov}}, \bibinfo {author} {\bibfnamefont {H.}~\bibnamefont
  {J{\'{o}}nsson}}, \ and\ \bibinfo {author} {\bibfnamefont {V.~M.}\
  \bibnamefont {Uzdin}},\ }\href {\doibase 10.1103/PhysRevB.94.174418}
  {\bibfield  {journal} {\bibinfo  {journal} {Physical Review B - Condensed
  Matter and Materials Physics}\ }\textbf {\bibinfo {volume} {94}},\ \bibinfo
  {pages} {174418} (\bibinfo {year} {2016})}\BibitemShut {NoStop}%
\bibitem [{\citenamefont {Muhlbauer}\ \emph {et~al.}(2009)\citenamefont
  {Muhlbauer}, \citenamefont {Binz}, \citenamefont {Jonietz}, \citenamefont
  {Pfleiderer}, \citenamefont {Rosch}, \citenamefont {Neubauer}, \citenamefont
  {Georgii},\ and\ \citenamefont {Boni}}]{Muhlbauer2009}%
  \BibitemOpen
  \bibfield  {author} {\bibinfo {author} {\bibfnamefont {S.}~\bibnamefont
  {Muhlbauer}}, \bibinfo {author} {\bibfnamefont {B.}~\bibnamefont {Binz}},
  \bibinfo {author} {\bibfnamefont {F.}~\bibnamefont {Jonietz}}, \bibinfo
  {author} {\bibfnamefont {C.}~\bibnamefont {Pfleiderer}}, \bibinfo {author}
  {\bibfnamefont {A.}~\bibnamefont {Rosch}}, \bibinfo {author} {\bibfnamefont
  {A.}~\bibnamefont {Neubauer}}, \bibinfo {author} {\bibfnamefont
  {R.}~\bibnamefont {Georgii}}, \ and\ \bibinfo {author} {\bibfnamefont
  {P.}~\bibnamefont {Boni}},\ }\href {\doibase 10.1126/science.1166767}
  {\bibfield  {journal} {\bibinfo  {journal} {Science}\ }\textbf {\bibinfo
  {volume} {323}},\ \bibinfo {pages} {915} (\bibinfo {year}
  {2009})}\BibitemShut {NoStop}%
\bibitem [{\citenamefont {Piacente}\ \emph {et~al.}(2010)\citenamefont
  {Piacente}, \citenamefont {Hai},\ and\ \citenamefont
  {Peeters}}]{Piacente2010}%
  \BibitemOpen
  \bibfield  {author} {\bibinfo {author} {\bibfnamefont {G.}~\bibnamefont
  {Piacente}}, \bibinfo {author} {\bibfnamefont {G.~Q.}\ \bibnamefont {Hai}}, \
  and\ \bibinfo {author} {\bibfnamefont {F.~M.}\ \bibnamefont {Peeters}},\
  }\href {\doibase 10.1103/PhysRevB.81.024108} {\bibfield  {journal} {\bibinfo
  {journal} {Physical Review B}\ }\textbf {\bibinfo {volume} {81}},\ \bibinfo
  {pages} {024108} (\bibinfo {year} {2010})}\BibitemShut {NoStop}%
\bibitem [{\citenamefont {Karapetrov}\ \emph {et~al.}(2009)\citenamefont
  {Karapetrov}, \citenamefont {Milo{\v{s}}evi{\'{c}}}, \citenamefont
  {Iavarone}, \citenamefont {Fedor}, \citenamefont {Belkin}, \citenamefont
  {Novosad},\ and\ \citenamefont {Peeters}}]{Karapetrov2009}%
  \BibitemOpen
  \bibfield  {author} {\bibinfo {author} {\bibfnamefont {G.}~\bibnamefont
  {Karapetrov}}, \bibinfo {author} {\bibfnamefont {M.~V.}\ \bibnamefont
  {Milo{\v{s}}evi{\'{c}}}}, \bibinfo {author} {\bibfnamefont {M.}~\bibnamefont
  {Iavarone}}, \bibinfo {author} {\bibfnamefont {J.}~\bibnamefont {Fedor}},
  \bibinfo {author} {\bibfnamefont {A.}~\bibnamefont {Belkin}}, \bibinfo
  {author} {\bibfnamefont {V.}~\bibnamefont {Novosad}}, \ and\ \bibinfo
  {author} {\bibfnamefont {F.~M.}\ \bibnamefont {Peeters}},\ }\href {\doibase
  10.1103/PhysRevB.80.180506} {\bibfield  {journal} {\bibinfo  {journal}
  {Physical Review B}\ }\textbf {\bibinfo {volume} {80}},\ \bibinfo {pages}
  {180506} (\bibinfo {year} {2009})}\BibitemShut {NoStop}%
\end{thebibliography}%

\end{document}